\documentstyle[12pt,a4]{article}

\begin{document}

\textheight 21.7 true cm
\textwidth 14.1 true cm

\voffset -0.7 true cm
\hoffset 0.5 true cm

\baselineskip 16pt
\parskip 8pt
\parindent 28pt

\title{`Thermal' ambience and fluctuations in classical field theory} 
\author{K.~Srinivasan\thanks{srini@iucaa.ernet.in}\, , \quad 
L.~Sriramkumar\thanks{lsk@iucaa.ernet.in} \quad and \quad T.~Padmanabhan
\thanks{paddy@iucaa.ernet.in}\\
IUCAA, Post Bag 4, Ganeshkhind, Pune 411 007, INDIA.}

\maketitle

\begin{abstract}
A plane monochromatic wave will not appear monochromatic to a noninertial
observer. 
We show that this feature leads to a `thermal' ambience in an 
accelerated frame {\it even in classical field theory}. 
When a real, monochromatic, mode of a scalar field is Fourier analyzed 
with respect to the proper time of a uniformly accelerating observer, 
the resulting power spectrum consists of three terms: (i)~a factor $(1/2)$ 
that is typical of the ground state energy of a quantum oscillator, 
(ii)~a Planckian distribution $N(\Omega)$ and---most 
importantly---(iii)~a term $\sqrt{N(N+1)}$, which is the root 
mean square fluctuations about the Planckian distribution.  
It is the appearance of the root mean square fluctuations that motivates 
us to attribute a `thermal' nature to the power spectrum. 
This result shows that some of the `purely' quantum mechanical results
might have a classical analogue.
The `thermal' ambience that we report here also proves to be a feature 
of observers stationed at a constant radius in the Schwarzschild and 
de-Sitter spacetimes. 
\end{abstract} 
\newpage

\section{Introduction}

It is well known that quantization of a field in Minkowski and Rindler
coordinates are not equivalent~\cite{fulling73}.
It is also known that the response of a uniformly accelerating detector 
in the Minkowski vacuum is a thermal spectrum~\cite{unruh76,dewitt79}. 
In both these situations, one obtains the thermal spectrum in the strict 
sense of the word: Not only that the mean occupation number in any mode 
is Planckian, but the fluctuations around the mean is also characterized 
by the standard thermal noise. 
These results suggest that the quantum fluctuations in the vacuum appear 
as thermal fluctuations in the uniformly accelerated frame.
Similar results also arise in Schwarzschild and 
de-Sitter spacetimes~\cite{hawking75,gibbons77}.

In contrast to quantum theory, classical field theory does not admit any
intrinsic fluctuations.  
The absence of concepts such as vacuum and fluctuations in classical 
field theory may lead us to believe that non-trivial phenomena as the 
one mentioned above will not have any classical analogue. 
We shall show, however, that such is not the case.
In this paper, we discuss a fairly non-trivial and interesting effect that 
arises purely in the context of {\it classical} field theory and which 
probably has serious implications for such phenomena as black hole entropy. 
We find that, when a {\it real} monochromatic wave mode of a classical 
field is Fourier transformed with respect to the proper time of a 
uniformly accelerating observer, the resulting power spectrum has a 
`thermal' nature---both as regards the mean occupation number {\it 
as well as the fluctuations around the mean}. 
Similar results arise in the Schwarzschild and de Sitter spacetimes 
as well.  

This paper is organized as follows.  
In sections~\ref{sec:acc} and~\ref{sec:gen} we show that a `thermal'
ambience is a feature of uniformly accelerated observers in Minkowski 
spacetime. 
In section~\ref{sec:bhdes} we explain as to how such an effect can 
arise in Schwarzschild and de-Sitter spacetimes too.
Finally, in section~\ref{sec:concl} we discuss the possible implications 
of our analysis. 

\section{`Thermal' ambience in an accelerated frame}\label{sec:acc}

Consider a massless, minimally coupled, scalar field which satisfies 
the Klein-Gordon equation
\begin{equation}
\Box\Phi\equiv{\Phi^{;\mu}}_{;\mu}=0.\label{eqn:kg}
\end{equation}
In flat spacetime, the basis solutions to the above Klein-Gordon equation 
in the Minkowski coordinates~$(t, {\bf x})$ can be taken to be plane waves
labeled by the wave vector ${\bf k}$: 
\begin{equation}
\Phi(t, {\bf x}) =  \cos \,(\omega t-{\bf k}.{\bf x}),\label{eqn:realwave}
\end{equation}
where $\omega=\vert {\bf k}\vert$. 
We now ask: Consider an observer who is moving on an arbitrary trajectory
$(t({\tau}), {\bf x}({\tau}))$, parametrized by the proper time ${\tau}$. 
How will this observer view the above Minkowski plane wave mode?

The moving observer will see the scalar field varying with respect to
his (her) proper time in a manner determined by the function  
$\Phi\left[t({\tau}), {\bf x}({\tau})\right]$.
If the observer is in inertial motion  then the monochromatic wave will
appear to be another monochromatic wave with a Doppler shifted frequency. 
But in general, for noninertial trajectories, the wave will not appear
to be monochromatic for the moving observer but will prove to be a
superposition of waves with different frequencies. 
To determine the exact decomposition of the wave, we should Fourier 
analyze the Minkowski mode in the frame of the observer. 
The Fourier transform of the Minkowski plane wave with respect to the 
proper time ${\tau}$ of the observer in motion is described by the 
integral
\begin{equation}
\tilde{\Phi}(\Omega)= \int_{-\infty}^{\infty} d\tau\; 
\Phi\left[t(\tau), {\bf x}(\tau)\right]\, 
e^{-i \Omega \tau}.\label{eqn:ftrans} 
\end{equation}
This expression gives the amplitude of a component with frequency $\Omega$
(as defined by the moving observer) present in the original monochromatic
wave.
Given a particular plane wave, we can always align the coordinates such 
that the wave is traveling along the $x$ axis, {\it i.e.} the wave vector 
is given by ${\bf k}=(k, 0, 0)$. 
Then the plane wave mode~(\ref{eqn:realwave}) reduces to
\begin{equation}
\Phi(t, {\bf x}) =  \cos(\omega t- k x)\label{eqn:rw}
\end{equation}
and its Fourier transform is given by the integral
\begin{equation}
\tilde{\Phi}(\Omega) 
= \int_{-\infty}^{\infty} d\tau\;
\cos\left[\omega t(\tau)- k x(\tau)\right]\;
e^{-i \Omega \tau}\;.\label{eqn:rwftrans}
\end{equation}

We shall now specialize to the case of an observer who is accelerating 
uniformly with respect to the Minkowski coordinates with a proper 
acceleration $g$.
We shall also assume that the observer is accelerating along the $x$-axis.  
The world line of such an observer in the Minkowski coordinates
$(t,x,y,z)$ is given by the relations~\cite{rindler66}
\begin{equation} 
t = t_0 + g^{-1} \,\sinh(g\tau)\quad ; \quad 
x = x_0 + g^{-1} \,\cosh(g\tau)\quad ; \quad 
y=y\quad {\rm and} \quad z=z,\label{eqn:rindcoords} 
\end{equation}
where $t_0$ and $x_0$ are constants and $\tau$ is the proper time as measured
by the clock in the frame of the uniformly accelerated observer.
The world line of such a uniformly accelerating observer is a hyperbola 
in the $(t, x)$ plane parametrized by the two constants $t_0$ and $x_0$.
The asymptotes of this hyperbola are the past and the future light
cones that intersect at the point $(t_0, x_0)$.
To see how the plane wave~(\ref{eqn:realwave}) will be viewed by such an 
observer, we substitute the coordinate transformations~(\ref{eqn:rindcoords}) 
in the Fourier integral~(\ref{eqn:rwftrans}), and obtain that~\cite{gandr80}
\begin{eqnarray}
\tilde{\Phi}(\Omega)
&=&\int_{-\infty}^{\infty} d\tau\,
\cos\biggl(\omega [t_0 -x_0 + g^{-1} \sinh(g\tau)
- g^{-1} \cosh(g\tau)]\biggl)\; 
e^{-i\Omega {\tau}}\nonumber\\
&=& \int_{-\infty}^{\infty} d\tau\,
\cos\left(\omega g^{-1} e^{-g \tau}
-\beta\right)\;e^{-i\Omega  \tau}\nonumber\\
&=& \left({1 \over 2g}\right)\, e^{-i\phi}\;
\left(e^{-(\Omega /4\Omega_0)}\, e^{-i\beta} 
+ e^{(\Omega/4\Omega_0)}\,e^{i\beta}\right)\;
\Gamma\left(i\Omega g^{-1} \right),
\end{eqnarray}
where
\begin{equation}
\phi = \Omega g^{-1}\; \ln(\omega g^{-1})
\qquad ; \qquad
\Omega_0=g/2\pi
\qquad {\rm and}\qquad
\beta =\omega(t_0-x_0).\label{eqn:phibeta}
\end{equation}
In the above integral we have assumed that the plane wave is traveling 
to the right so that $k=\omega$.
The resulting power spectrum per logarithmic interval in frequency is given 
by ${\cal P}(\Omega)\equiv \left(\Omega\,{\vert{\tilde \Phi}(\Omega)
\vert}^2\right)$ and can be written in a remarkable form:
\begin{eqnarray}
{\cal P}(\Omega)\equiv \Omega\; {\vert{\tilde \Phi}(\Omega)\vert}^2
&=& \left({\pi \over 2g}\right)\; 
\biggl(\coth\left(\Omega/2\Omega_0\right)
+ {\rm csch}\left(\Omega/2\Omega_0\right)\; 
\cos(2\beta)\biggl)\nonumber\\
&=& \left({\pi\over g}\right)\;
\; \left\lbrace{1 \over 2} + N
+ \sqrt{N(N+1)}\; \cos(2\beta)\right\rbrace, \label{eqn:ps}
\end{eqnarray}
where 
\begin{equation}
N(\Omega)=\left({1 \over 
{\exp\left(\Omega/\Omega_0\right)-1}}\right).\label{eqn:planck}
\end{equation}
We shall now consider various features of this result.

To begin with we note that this result is a purely classical one and 
hence $\hbar$ does not appear anywhere. 
In ordinary units, $\Omega_0=(g/2\pi c)$ has the correct dimensions, 
{\it viz.} per second, for a frequency.
The quantity $N(\Omega)$ is a Planckian in terms of {\it frequencies} and 
is again independent of $\hbar$. 
Usually, one tries to  express the Planckian distribution in terms 
of energies of the `quanta' labeled by frequency $\Omega$ and in
such a case we need to  write frequencies as, say $\Omega=(E/\hbar)$,
thereby {\it artificially} introducing $\hbar$; but the result, stated 
as a power spectrum in frequency space, makes perfect conceptual sense 
as it stands.
(For example, radio astronomers measure the power spectrum in frequency
space and may not think in terms of photons.)
Of course, to obtain a quantity with the dimension of temperature we 
again need to introduce a $\hbar$ into the quantity $\Omega_0$.
(For this reason the word thermal has been appearing within quotes 
throughout this paper.)

The analysis done above could have been carried out even in the days
before quantum theory---it uses only classical relativity.
But it is our knowledge of quantum theory that allows a suggestive 
interpretation of the three terms in the power spectrum: 
The first term---{\it viz.} the  factor $(1/2)$---is typical of the ground 
state energy of a quantum oscillator.
The second term $N$ is a Planckian distribution in $\Omega$, as already
mentioned.
{\it Note that these two terms are totally independent of the original 
frequency $\omega$ of the plane wave!} 

The third term is still more remarkable.
When we vary the constants $t_0$ and $x_0$ this term fluctuates between
$-\sqrt{N(N+1)}$ and $+\sqrt{N(N+1)}$. 
The magnitude of this fluctuation (which is the root mean square 
deviation about the mean value) is exactly what one would have obtained 
for a strictly thermal distribution of massless bosonic quanta in quantum 
field theory. 
In fact it is this fluctuation that motivates us to attribute a `thermal'  
nature to the power spectrum.

While this result is very suggestive, it must be noted that $\beta$ is 
related to $t_0$ and $x_0$ by equation~(\ref{eqn:phibeta}).  If the original
plane wave had an extra phase ${\delta}$, then the argument of the cosine
term will pick up $2\delta$ additionally.
So by choosing the constants $\delta, t_0$ and $x_0$ suitably, it  
possible to kill the fluctuations in the power spectrum. 
(It is also easy to verify that one {\it cannot} choose the constants to 
cancel the first two terms as well.)
The implications of this result are not clear.

It may be noted that the existence of the three terms is a direct 
consequence of our choosing a {\it real} plane wave. 
If the same analysis is repeated for a complex mode for the 
scalar field, say $\Phi(t,x)=exp-i(\omega t-kx)$, then the 
resultant power spectrum per logarithmic frequency interval 
is given by
\begin{equation}
{\cal P}(\Omega)= \left({2\pi \over g}\right)\; N,
\end{equation}
where $N$ is given by~(\ref{eqn:planck}).
We do not get the zero-point term or the fluctuations. 
Of course, in classical field theory, one must use {\it real} modes 
and that is eaxactly what we have done here.

Finally, let us consider the limit of $\omega\to 0$. 
In this limit, the field in the inertial frame reduces to an unimportant 
constant---which could be thought of as closest to the concept of a 
`vacuum' in the classical theory. 
The Fourier integral as well as the phase $\phi$ in 
equation~(\ref{eqn:phibeta}) diverges when $\omega\to 0$; 
but the power spectrum---which is the squared modulus of 
the amplitude---is well defined:
\begin{equation}
{\cal P}(\Omega){\biggl\vert}_{\,\omega\to 0}=
\left({\pi\over g}\right)\;
\; \left\lbrace{1 \over 2} + N
+ \sqrt{N(N+1)}\;\right\rbrace.\label{eqn:pszeroomega}
\end{equation}
However, as long as $\omega$ is treated as a `regulator' one can say that 
the accelerated observer will see these terms even in the limit 
of $\omega\to 0$. 
This is very reminiscent of the inertial vacuum appearing as a Planckian 
spectrum to the accelerated observer in a manner which is completely
independent of the original wave mode.

\section{Generalization to other field configurations}\label{sec:gen}

In the last section we have carried out our analysis for real Minkowski 
waves that were traveling to the right.
It is straight forward to verify that one obtains the same power spectrum 
for left moving waves, {\it i.e} when $k=-\omega$.

A more general case is as follows. 
Consider a function of $\Phi(t-x)$ that satisfies the Klein-Gordon 
equation and  is either odd or even in $(t-x)$.
Such a function $\Phi(t-x)$, which will represent a wave packet that is 
traveling along the $x$ axis, can be Fourier decomposed into the following 
form
\begin{equation}
\Phi(t-x)= \int_{-\infty}^{\infty} d\alpha\,
f(\alpha)\, \exp {i\alpha (t-x)}.\label{eqn:ft}
\end{equation}
The function $f(\alpha)$ will prove to be odd or even depending on whether
$\Phi(t-x)$ is odd or even.
Substituting the transformation equations~(\ref{eqn:rindcoords}) 
in~(\ref{eqn:ft}) and Fourier transforming as before with respect 
to the proper time of the Rindler observer, we obtain that 
\begin{equation}
{\tilde \Phi}(\Omega) = g^{-1}\; 
\Gamma(i\Omega g^{-1})\; \left(e^{(\Omega/4\Omega_0)}\, F_1(\Omega)
\pm e^{-(\Omega/4\Omega_0)}\, F_2(\Omega)\right),\label{eqn:asoddeven}
\end{equation}
where the plus sign is to be chosen if $\Phi(t-x)$ is an even function
and the minus sign if $\Phi(x-t)$ is an odd function. 
The distributions $F_{1}(\Omega)$ and $F_{2}(\Omega)$ are described by 
the integrals
\begin{equation}
F_{1}(\Omega) = \int_{0}^{\infty} d\alpha\; f(\alpha)\; 
\exp-\left(i\Omega g^{-1}\ln(g^{-1}\alpha)\right)\, 
e^{i\alpha (t_0 - x_0)}
\end{equation}
and
\begin{equation}
F_{2} (\Omega) = \int_{0}^{\infty} d\alpha\; f(\alpha)\; 
\exp-\left(i\Omega g^{-1}\ln(g^{-1}\alpha)\right)\, 
e^{-i\alpha (t_0 - x_0)},
\end{equation}
where $\Omega_0$ is given by~(\ref{eqn:phibeta}). 
Now we obtain that
\begin{eqnarray}
{\cal P}(\Omega) &\equiv& 
\Omega\;{\vert {\tilde \Phi}(\Omega)\vert}^2\nonumber\\
&=& \left({\pi \over {g \sinh(\Omega/2\Omega_0)}}\right)\; 
\Biggl\lbrace e^{(\Omega/2\Omega_0)}\,\vert F_{1}(\Omega)\vert^2 
+ e^{-(\Omega/2\Omega_0)}\,\vert F_{2}(\Omega)\vert^2 \nonumber\\ 
& & \qquad\qquad\qquad\qquad\qquad\quad 
\pm \biggl(F_1^* (\Omega) F_2 (\Omega) 
+ F_{1} (\Omega)F_2^* (\Omega)\biggl)\Biggl\rbrace.\label{eqn:psoddeven}
\end{eqnarray}
This spectrum, of course, does not have a thermal nature 
since it depends explicitly on the form of $f(\alpha)$.  

But a simplification occurs if we treat $f(\alpha)$ as a stochastic 
variable so that when averaged over an ensemble of realizations, it 
satisfies the relation
\begin{equation}
\langle  f(\alpha)\, f^{*}({\alpha'})\rangle 
= P(\alpha)\;\delta({\alpha -\alpha'}),
\end{equation}
with some power spectrum $P(\alpha)$, such that 
$\int_{-\infty}^{\infty}d\omega\, P(\omega)=2C$. 
In such a case, when $\vert F_{1}(\Omega)\vert^2$ and 
$\vert F_{2}(\Omega)\vert^2$ are averaged over the stochastic 
variable $f(\alpha)$, both reduce to a constant independent 
of $\Omega$, {\it i.e.} 
\begin{equation}
\langle {\vert F_{1}(\Omega)\vert}^2\rangle 
= \langle {\vert F_{2}(\Omega)\vert}^2\rangle 
= \int_{0}^{\infty} d\alpha\, P(\alpha) = C.
\end{equation}
The power spectrum~(\ref{eqn:psoddeven}) when averaged over 
the stochastic variable $f(\alpha)$ is given by
\begin{equation}
\langle {\cal P}(\Omega)\rangle 
= \left({4\pi C\over g}\right) 
\left\lbrace{1 \over 2} 
+ N \pm \sqrt{N(N+1)}\,\cos(2\beta')\right\rbrace,
\end{equation}
where $\beta'$ is a function of $(t_0 - x_0)$ and is defined by the 
relation
\begin{eqnarray}
\cos(2\beta') &=& \left({1 \over 2C}\right)\;
\biggl\langle F_1^* (\Omega)F_2 (\Omega) 
+ F_1 (\Omega)F_2^* (\Omega)\biggl\rangle\nonumber\\
&=& \left({1 \over C}\right)\, \int_{0}^{\infty} d\alpha\; P(\alpha)\,
\cos[2\alpha(t_{0}-x_{0})].
\end{eqnarray}  
So a stochastic wave field in the Minkowski frame will also reproduce all 
the three terms in the power spectrum obtained earlier.

The wave field described above did not have explicit random phases. 
It is possible to define a random wave field in a different way.
Consider the following random superposition of real modes for the scalar 
field 
\begin{equation}
\Phi(t,x)=\int_{-\infty}^{\infty} d\omega\, A(\omega)\,
\cos\left[\omega(t-x)+\theta(\omega)\right],\label{eqn:randomrw}
\end{equation}
where $A(\omega)$ and $\theta(\omega)$ are stochastic variables 
satisfying the relations
\begin{equation}
\langle A(\omega)\,A(\omega')\rangle 
= {\bar P}(\omega)\;\delta({\omega-\omega'}) 
\qquad ; \qquad
\langle \theta(\omega)\rangle = 0
\end{equation}
and ${\bar P}(\omega)$ is an arbitrary function of $\omega$ such that   
${\bar C}=\int_{-\infty}^{\infty}d\omega\, {\bar P}(\omega)$ is a 
finite constant.  
We can now set $t_0=x_0=0$ in~(\ref{eqn:rindcoords}) without any 
loss of generality.
Substituting the coordinate transformations~(\ref{eqn:rindcoords}) 
in the scalar field configuration given by~(\ref{eqn:randomrw})
and Fourier transforming the same with respect to the proper time 
of the uniformly accelerated observer, we obtain
\begin{eqnarray}
\tilde{\Phi}(\Omega)&=& \int_{-\infty}^{\infty} d\tau\, 
\int_{-\infty}^{\infty} d\omega\, A(\omega)\, 
\cos\biggl(\omega\left[t(\tau)-x(\tau)\right]
+\theta(\omega)\biggl)\; e^{-i \Omega \tau}\nonumber\\ 
&=& \int_{-\infty}^{\infty} d\omega\, A(\omega)\,
\int_{-\infty}^{\infty} d\tau \, 
\cos\biggl(\omega g^{-1} e^{-g\tau} 
- \theta(\omega)\biggl)\; e^{-i\Omega \tau}\nonumber\\
&=&\left({1\over 2g}\right)\; \Gamma(i\Omega g^{-1}) 
\int_{-\infty}^{\infty} d\omega\; 
A(\omega)\;e^{-i\phi}\nonumber\\
& & \qquad\qquad\qquad\quad\quad\quad
 \times \left(e^{-(\Omega/4\Omega_0)}\, e^{-i\theta(\omega)} 
+ e^{(\Omega/4\Omega_0)}\,e^{i\theta(\omega)}\right),
\end{eqnarray}
where ${\phi}$ and $\Omega_0$ are given by~(\ref{eqn:phibeta}).
The power spectrum per logarithmic frequency interval, {\it viz.} the 
quantity $\left(\Omega\, {\vert \tilde{\Phi}(\Omega)\vert}^2\right)$ 
when averaged over the stochastic variables $A(\omega)$ and 
$\theta(\omega)$ then reduces to  
\begin{equation}
\langle {\cal P}(\Omega)\rangle 
=  \left({\pi {\bar C}\over g}\right)\,
\left\lbrace{1 \over 2} + N \;\right\rbrace.\label{eqn:randps}
\end{equation}

In this case, the random phases have averaged out the fluctuation
term, {\it viz.} the factor $\sqrt{N(N+1)}$ that had appeared in 
the power spectrum~(\ref{eqn:ps}).
A somewhat similar result was obtained earlier by Boyer~\cite{boyer80}.
He modeled the zero-point fluctuations as due to
random superposition of Minkowski
plane wave modes, and used it as a basis for investigating the `spectrum'  
observed by a uniformly accelerating observer.
He showed that the correlation function of an accelerating observer 
`in a random classical scalar zero-point radiation' exactly matches 
the correlation function of an inertial observer in a thermal background. 
Our analysis here shows that the effect reported by Boyer 
arises when a random superposition of Minkowksi real modes 
are simply Fourier analyzed in the frame of a uniformly 
accelerating observer~(cf. equation~(\ref{eqn:randps})). 
{\it But notice that, such an approach has killed a very interesting 
$\sqrt{N(N+1)}$ term which was originally present.}

Finally, we discuss a case in which the observer is moving in a direction
perpendicular to the wave vector.
Consider an observer who is uniformly accelerating along the $y$ axis, 
{\it i.e.} in a direction perpendicular to which the plane wave is 
traveling (which we always take to be the $x$-axis).
If the proper acceleration of the observer is $g$, then the coordinate 
transformations to the uniformly accelerated frame are given by
\begin{equation}
t = t_0 + g^{-1} \sinh(g\tau)\quad ;\quad
x=x\quad ;\quad
y = y_0 + g^{-1}  \cosh(g\tau)\quad {\rm and}\quad
z=z.
\end{equation}
Substituting these transformations in the Fourier 
transform~(\ref{eqn:rwftrans}), we obtain
\begin{eqnarray}
{\tilde \Phi}(\Omega)
&=& \int_{-\infty}^{\infty} d\tau\, 
\cos\left(\omega (t_0 + g^{-1}  \sinh(g\tau))- k x\right)\; 
e^{-i\Omega \tau}\nonumber\\
&=& g^{-1}\,
K_{(i\Omega/ g)}\left(\omega g^{-1}\right)\,
\left(e^{-(\Omega/4\Omega_0)}\, e^{-i(\omega t_0-kx)}
+ e^{(\Omega/4\Omega_0)}\, e^{i(\omega t_0-kx)}\right),
\end{eqnarray}
where $K_{(i\Omega/g)}$ is the Bessel function of imaginary order.
The resulting power spectrum 
\begin{eqnarray}
{\cal P}(\Omega)
&\equiv& \Omega\;{\vert {\tilde \Phi}(\Omega)\vert}^2\nonumber\\
&=& 2\Omega g^{-2}\,
{\biggl\vert K_{(i\Omega/g)}\left(\omega g^{-1}\right)\biggl\vert}^2\;
\biggl\lbrace\cosh(\Omega/2\Omega_0) 
+ \cos[2(\omega t_0-kx)]\biggl\rbrace\nonumber\\
&=& 4\Omega g^{-2}\,\sinh(\Omega/2\Omega_0)\; 
{\biggl\vert K_{(i\Omega/g)}
\left(\omega g^{-1}\right)\biggl\vert}^2\nonumber\\ 
& & \qquad\qquad\quad\quad
\times\;\left\lbrace{1 \over 2} + N 
+ \sqrt{N(N+1)}\,\cos[2(\omega t_0-kx)]\right\rbrace,\label{eqn:psy}
\end{eqnarray}
does not have a thermal nature because of the coefficients 
multiplying the expression in the curly brackets.
Therefore, thermal ambience arises only for observers whose
acceleration is along the same axis as the direction of 
propagation of the wave. 

It is however interesting to ask: What happens to the power 
spectrum~(\ref{eqn:psy}) in the limit of $\omega\to 0$?
In this limit, the original wave field is a constant and any 
direction of motion for the observer should be equivalent. 
Hence we expect to see the `thermal' ambience in this limit 
even for this observer. 
This is indeed the case:
In the limit of ${\omega\to 0}$  
\begin{equation}
K_{(i\Omega g^{-1})}\left(\omega g^{-1}\right) 
\approx 2^{(i\Omega g^{-1}-1)}\;(\omega g^{-1})^{-(i\Omega g^{-1})}
\;\Gamma(i\Omega g^{-1}).
\label{eqn:approx}
\end{equation}
Substituting the above  approximation for $K_{(i\Omega g^{-1})}
\left(\omega g^{-1}\right)$ in (\ref{eqn:psy}) one recovers the 
result given in~(\ref{eqn:ps}) with $\beta$ set to zero.
This result also holds for a wave propagating in an arbitrary 
direction, as is to be expected.

\section{`Thermal' ambience in Schwarzschild and de-Sitter 
spacetimes}\label{sec:bhdes}

In this section, we shall briefly comment on the generalization of 
the above results to Schwarzschild and de-Sitter spacetimes. 
The solution to the Klein-Gordon equation in these spacetimes cannot 
be expressed in terms of simple functions in (3+1) dimensions and 
hence we will work in (1+1) dimensions.
  
In (1+1)~dimensions, the Schwarzschild spacetime is described by the 
line-element
\begin{equation}
ds^2= \left(1-{2M \over r}\right)dt^2
- {\left(1- {2M\over r}\right)}^{-1}\,dr^2.
\end{equation}
In terms of the Regge-Wheeler coordinates $(t, r^*)$~\cite{mtw}, where
\begin{equation}
r^*=r+2M\,\ln\left({{r \over 2M}-1}\right),
\end{equation}
the Schwarzschild line-element turns out to be conformal to the flat 
space metric, {\it i.e.}
\begin{equation}
ds^2 = \left(1 - {2M \over r}\right)\, (dt^2 - {dr^*}^2).
\end{equation}
And, in terms of the Kruskal-Szekeres coordinates $(v, u)$~\cite{mtw}, 
which are related to the Regge-Wheeler coordinates $(t, r^*)$ by the 
transformations
\begin{equation}
u = u_0 + e^{r^*/4M}\, \cosh(t/4M)\qquad {\rm and}
\qquad v = v_0 + e^{r^*/4M}\, \sinh(t/4 M)\label{eqn:uvs},
\end{equation}
(where $u_0$ and $v_0$ are arbitrary constants) the Schwarzschild 
line-element reduces to
\begin{equation}
ds^2 = \left({32 M^3 \over r}\right)\, e^{-(r/2 M)}\; (dv^2 - du^2). 
\end{equation}
The proper time $\tau$ of an observer stationed at a constant $r$ is 
then related to the Schwarzschild time coordinate $t$ by the equation 
\begin{equation}
\tau=\lambda(r)\,t
\qquad {\rm where}\qquad 
\lambda(r)={\left(1-{2M \over r}\right)}^{1/2}.
\end{equation}
Just as the trajectory of a uniformly accelerating observer is a
hyperbola in the plane of the Minkowski coordinates, the world line
of an observer stationed at a constant $r$ is a hyperbola in the 
$(v, u)$ plane.
And, the asymptotes of this hyperbola are the past and the future 
horizons of the Schwarzschild spacetime that intersect at the point 
$(v_0, u_0)$.

As is well known the action for a minimally coupled scalar field
is conformally invariant in (1+1) dimensions. 
Hence the normal modes of a massless, minimally coupled scalar field 
in conformally flat metrics are just plane waves.
So the normal mode solutions of the Schwarzschild spacetime in the
Kruskal-Szekeres coordinates $(v, u)$ are just plane waves. 
Consider a single real mode described by the equation
\begin{equation}
\Phi(v, u) = \cos\,(\omega v - ku).\label{eqn:kspw}
\end{equation}
We would like to know how an observer located at constant (Schwarzschild)
radial coordinate $r$ will describe this mode.
Assuming that the plane wave is traveling to the right, {\it i.e.}
$k=\omega$ and Fourier tranforming the monochromatic wave given in
equation~(\ref{eqn:kspw}) with respect to the proper time $\tau$ of an
observer stationed at a constant $r$, we obtain that
\begin{eqnarray}
\tilde\Phi(\Omega) &=& \int_{-\infty}^{\infty} d\tau\;
\Phi[v(\tau), u(\tau)]\; e^{-i\Omega \tau}\nonumber\\
&=& \lambda\, \int_{-\infty}^{\infty} dt\,
\cos \left(\omega e^{(r^*-t)/4M} - \beta\right)\; 
e^{-i\Omega \lambda t}\nonumber\\
&=& 2M\lambda\; e^{-i\mu}\;
\left(e^{-2\pi \Omega M \lambda}\,e^{-i\beta} 
+ e^{2\pi \Omega M \lambda}\,e^{i\beta}\right)\; 
\Gamma\left(4i\Omega M \lambda\right),
\end{eqnarray}
where 
\begin{equation}
\mu= 4\Omega M \lambda\; \ln\left(\omega e^{r^*/4M}\right) 
\qquad {\rm and}\qquad
\beta=\omega (v_0 - u_0).
\end{equation}\label{eqn:thetas}
The resulting power spectrum per logarithmic frequency interval is
then 
\begin{equation}
{\cal P}(\Omega)\equiv\Omega\;{\vert {\tilde \Phi}(\Omega)\vert}^2 
=\left(4 \pi M  \lambda\right)\; 
\left\lbrace{1 \over 2} + N+ \sqrt{N(N+1)}\, \cos(2\beta)\right\rbrace
\end{equation}
where
\begin{equation}
N(\Omega) 
= \left({1 \over {\exp\left(8\pi M \Omega \lambda\right)- 1}}\right).
\end{equation}
We once again obtain the three terms discussed before. 

The analysis for the de-Sitter spacetime is similar. 
The line-element that describes the de-Sitter spacetime is
\begin{equation}
ds^2 = (1-H^2r^2)\, dt^2 - {(1-H^2r^2)}^{-1}\,dr^2.
\end{equation}
In terms of the `Regge-Wheeler' coordinates $(t, r^*)$ corresponding
to the de-Sitter spacetime, where
\begin{equation}
r^* = H^{-1}\, {\rm arctanh}(Hr).\label{eqn:r*ds}
\end{equation}
the de-Sitter line-element turns out to be 
\begin{equation}
ds^2 = (1-H^2r^2)\,(dt^2 - {dr^*}^2).
\end{equation}
The `Kruskal-Szekeres' coordinates $(v, u)$ corresponding to the 
de-Sitter spacetime are related to the coordinates `Regge-Wheeler' 
coordinates $(t, r^*)$ by the equations
\begin{equation}
u = u_0 + e^{H r^*}\, \cosh(H t) \qquad {\rm and} \qquad 
v = v_0 + e^{H r^*}\, \sinh (H t).\label{eqn:uvds}
\end{equation}
The de-Sitter line-element in terms of the coordinates $(v, u)$ then 
reduces to
\begin{equation}
ds^2=H^{-2}\, (1-Hr)^2\, (dv^2-du^2). 
\end{equation}
Consider an observer who is stationed at a constant $r$ in de-Sitter
spacetime.
The world line of such an observer, just as in the Schwarzschild case, 
is a hyperbola in the $(v, u)$ plane whose asymptotes are the past and 
the future horizons of the de-Sitter spacetime that intersect at the 
point $(v_0, u_0)$. 
The proper time $\tau$ of this observer is related to the de-Sitter time
coordinate $t$ as follows
\begin{equation}
\tau=\lambda\, t,
\qquad{\rm where \;\;\; now}\qquad
\lambda= {(1-H^2r^2)}^{1/2}.
\end{equation}
For the case of a real wave as given in~(\ref{eqn:kspw}), 
where the coordinates $v$ and $u$ are now related to de-Sitter 
coordinates $t$ and $r$ by the equations~(\ref{eqn:uvds}) 
and~(\ref{eqn:r*ds}), the power spectrum per logarithmic frequency
interval as seen by the observer 
stationed at a constant $r$ is
\begin{equation}
{\cal P}(\Omega) \equiv\Omega\,{\vert {\tilde \Phi}(\Omega)\vert}^2
= (\pi  H^{-1}\lambda)\;
\left\lbrace{1 \over 2} + N + \sqrt{N(N+1)}\, \cos(2\beta)\right\rbrace,
\end{equation}
where
\begin{equation}
N(\Omega)=\left({1 \over {\exp\left(2\pi \Omega H^{-1}\lambda\right)
-1}}\right).
\end{equation}
In evaluating the powers spectrum above, it has been assumed that 
$k=\omega$, so that $\beta=\omega (v_0 - u_0)$.
The similarity to the previous results are obvious.

\section{Conclusions}\label{sec:concl}

In conclusion, we would like to stress those aspects of our results 
which are unexpected and contrast them with those which could have 
been anticipated with some hindsight.

To begin with, the following fact is well-known: In quantum field theory, 
the amplitude for transition of an Unruh-DeWitt detector, up to the first 
order in perturbation theory, is described by an integral that is similar 
in form to~(\ref{eqn:ftrans})~\cite{unruh76,dewitt79}.  
When the scalar field is decomposed in terms of the Minkowski modes, the 
transition probability, per unit proper time, of a uniformly accelerating 
Unruh-DeWitt detector turns out to be a thermal spectrum (see for
instance~\cite{bandd82}).
It might, therefore, seem that when a traveling wave is 
Fourier transformed with respect to the proper time of a uniformly 
accelerated observer, the resulting power spectrum will have a thermal 
nature.

However, there are some subtilities involved. 
To begin with, the modes of the quantum field are complex while here 
we are dealing with real plane wave modes. 
This makes the vital difference. 
As we have mentioned before, while a complex mode like 
$exp-i(\omega t-kx)$ will give a Planckian distribution 
it will {\it not} yield the two other terms we have obtained 
in our analysis.
In this sense, the real wave is quite different from the complex one.
We stress the fact that, when a real Minkowski mode is Fourier transformed 
with respect to the proper time of a uniformly accelerating observer, 
the resulting power spectrum not only contains a Planckian distribution 
but also contains the root mean square fluctuations about the Planckian. 
As mentioned earlier, it is the appearance of these fluctuations that 
motivates us to attribute a `thermal' nature to the power spectrum. 
{\it We know of no simple way to guess at this answer}.

Secondly, note the effect survives in the power spectrum
even in the limit of ${\omega\to 0}$. 
This is the closest to what one can call a `classical' 
vacuum---and our result shows that such a mode, with infinitesimal 
frequency, leads to a thermal ambience in the accelerated frame 
which is {\it totally independent of the properties of the original 
wave}. 
This result suggest that there is a deep connection between plane 
waves, accelerated frames and thermal fluctuations even at the 
classical level. 
This connection could be worth exploring.

A somewhat similar analysis, {\it viz.} Fourier analyzing the 
Minkowski modes in the frame  of an uniformly accelerated observer 
was carried out earlier by Gerlach~\cite{gerlach88}.
He had constructed a linear superposition of Minkowski modes in
(3+1)~dimensions such that the modulus square of the amplitude of 
these modes (which represents the total classical energy of these 
modes) to be equivalent to that of the ground state energy of a 
quantum oscillator.
Fourier analyzing such a field configuration with respect to the proper 
time of a uniformly accelerating observer, Gerlach had obtained a power 
spectrum (in a particular semiclassical limit) similar in form to 
equation~(\ref{eqn:ps}). 
He had presented his result as a `heuristic derivation of the thermal 
spectrum' that arises in quantum field theory due to the inequivalent
quantization in Minkowski and Rindler coordinates.
Our results and emphasis are different in several ways. 
To begin with, the effect we are reporting here is a feature of classical 
field theory and no quantum processes are involved. It is physically
motivated in a clear and simple manner and we do not have to resort
to any superposition of modes.
Secondly, our results are {\it exact} for a real, monochromatic plane wave 
while Gerlach needed to resort to some approximations because of the 
particular superposition of modes he had chosen. 
Thirdly, we would like to draw attention to  the zero-frequency limit 
of the wave, when it takes a life of its own in the accelerated frame.
This result, as far as we know, has not been noted in the literature
before. 
Finally, Gerlach had offered no explanation for the appearance of the 
factor $\cos(2\beta)$ as the coefficient of the fluctuation term.
Our analysis clearly shows that it arises due to the shift in the 
origin of the Minkowski coordinates. 

In section~\ref{sec:bhdes} we have shown that a thermal ambience is a 
feature of black hole spacetimes too. 
There has been an earlier attempt by Frolov~\cite{frolov91} in which 
he had modeled the black hole as a black body cavity and obtained a 
thermal spectrum for the radiation leaking out of a cavity
But Frolov had invoked quantum theory to obtain his results. 
Now, knowing that a `thermal' ambience can be a feature of black hole 
spacetimes too, we are presently investigating the possibility of 
interpreting the origin of black hole entropy purely classically.

\section*{\centerline {Acknowledgments}}

\noindent
KS and LSK are being supported by Junior and the Senior research fellowships 
of the Council of Scientific and Industrial Research, India, respectively.

\end{document}